\documentclass[twoside]{irmaems}
\usepackage{amssymb} 
\usepackage{amsmath} 
\usepackage{latexsym}

\renewcommand\theequation{\arabic{section}.\arabic{equation}}

\theoremstyle{definition} 

 \newtheorem{definition}{Definition}[section]

\theoremstyle{plain}      

 \newtheorem{theorem}[definition]{Theorem}

\newcommand{\tr}{\operatorname{tr}}

\usepackage{color}
\usepackage{hyperref}

\def\bea{\begin{align}}\def\eea{\end{align}}
\def\ie{{\rm i.e.\/}\ }

\def\be{\begin{equation}}\def\ee{\end{equation}}
\def\bei{\begin{itemize}}\def\eei{\end{itemize}}

\newcommand{\nco}{\newcommand}
\nco{\one}{\ensuremath{\,\,\mathrm{l}\!\!\!1}} 
\nco{\NN}{\mathbb{N}}
\nco{\ZZ}{\mathbb{Z}}
\nco{\QQ}{\mathbb{Q}}
\nco{\RR}{\mathbb{R}}
\nco{\CC}{\mathbb{C}}
\nco{\HH}{\mathbb{H}}
\nco{\OO}{\mathbb{O}}
\nco{\red}{\color{red}}
\def\magenta#1{\color{magenta} #1}

\def\ommit#1{{}}
\def\oh{\frac{1}{2}}
\def\ie{{\it i.e.}\ }

\def\CL{{\cal L}}

\begin{document}

\title{Invariances in Physics and Group Theory}

\author{Jean-Bernard Zuber}

\address{
LPTHE, CNRS-UMR 7589 and Universit\'e Pierre et Marie Curie, \\
4 place Jussieu, 75252 
Paris Cedex 5, France\\
email:\,\tt{zuber@lpthe.jussieu.fr}}

\maketitle

\begin{abstract}
This is a short review of the heritage of Klein's Erlangen program in modern physics.\footnote{Talk given at the 
Conference {\it Lie and Klein; the Erlangen program and its impact on mathematics and physics},  Strasbourg, Sept. 2012}
\ommit{This is a ``sample'' freely taken from various articles in
IRMA Lectures in Mathematics and Theoretical Physics. The goal is to illustrate how
to work with the style macro called {\tt irmaems.cls}. This macro is based on the standard
\LaTeX 2e article style and icludes a slightly changed variant of amsthm.sty. Users who are familiar
with \LaTeX\ should have no problems to work with {\tt irmaems.cls} instead of {\tt article.cls}.
Please start your file with}

\end{abstract}



\tableofcontents   

\section{Introduction}\label{s-1}

Let us be honest: most physicists of our time, even theorists,  do not have a  very clear notion of
 what  Klein's {\it Erlanger Programm} is about, and this is an understatement\dots 
If we read Weyl \cite{Weyl28}, however, \\
 {``According to Klein's Erlanger Program any geometry of a point-field is based on a particular transformation 
 group $\mathfrak{G}$ of the field; figures which are equivalent with respect to $\mathfrak{G}$, and which can therefore
 be carried into one another by a transformation of $\mathfrak{G}$, are to be considered as the same\dots"}\\
  and substitute  ``physical systems" for ``figures", we see that modern 
physicists, like Moli\`ere's character Monsieur Jourdain, who was delighted to learn that he had
been speaking prose all his life without knowing, would love to hear that they keep following Klein's program\dots

The aim of this lecture is indeed to illustrate how group theory associated with invariances of the geometry or the 
dynamics of a physical system has pervaded all modern physics and has become of everyday use in the 
physicist's toolbox.  A word of caution, though.
As the author of these lines is not a professional historian of science, this lecture will undoubtedly present only a  
biased and incomplete view of this vast subject. 
 

\section{Early group theory in 19th century physics: crystallography}

Before the  birth of Lie group theory and Klein's Erlangen program, physicists had realized the role
of symmetry in Nature and foreseen the importance of group theory in physical sciences. 
 It had been known for  long  by artists that 2-dimensional periodic patterns 
-- tilings or wall paper motives, {\it i.e.} two-dimensional ``crystals'' -- were coming in finite types. There are 17 types of symmetry
-- 17 space groups in modern terminology -- in two dimensions. For a beautiful illustration, see the
web site \url{http://en.wikipedia.org/wiki/Wallpaper\_group}. Crystallographers then set out to tabulate the corresponding 
structures in  3 dimensions,
classifying in turn the point groups,  i.e. groups that fix a point of a lattice,   the classes of lattices, and the space 
groups, taking translations into account. This  long endeavour  kept them busy for the major part of the nineteenth century,
with important steps achieved by Frankenheim and by Hessel (32 point groups in 3 dimensions, in the 1830's), 
by Bravais (14 classes of lattices, circa 1850), C. Jordan (who emphasized the role of groups), and many others.
The program was completed in the  early nineties of that century,  by 
{Sch\"onflies\footnote{It may be relevant to observe that Sch\"onflies was directed to this problem by Klein, 
who saw it as a nice illustration of his program\dots}, Fedorov and Barlow} (1891--94), 
with the classification of the 230 space groups in 3 dimensions, see \cite{Burckhardt, Phillips, Janssen, Michel}. 
The situation is summarized in the 
following table

\begin{center}
\begin{tabular}{|c|c|c|c|}
\hline
{
Dimension} $d$ & Point Groups & Lattices &
{ Space groups }\\
\hline
$
{d=1}$ &2 & 1  & 
{7} \\
$d=2$ & 10& 5 & 
{17} \\
$d=3$ & 32 & 14  & 
 {230}\\
 \hline
 \end{tabular}\\
 \end{center} 
 
\medskip
According to H. Weyl \cite{Weyl28} 
{ { ``The most important application of group theory to natural science heretofore has been in this field"}. 
It is interesting to notice that Weyl wrote this comment in 1928, many years after the birth of Relativity -- both Special and General --,
and as he was himself working on the applications of group theory to quantum physics\dots

\smallskip
{\bf Breaking of symmetry.} If it is important in physical sciences to know the possible types of 
symmetry, it is maybe even more interesting to understand the way these symmetries may be broken
\footnote{This is also an interesting issue in art, see \cite{Weyl-sym}.}.
This was emphasized in a particularly clear way by
Pierre Curie, as stated in his  principle (1894) \cite{Curie}: ``{Elements of symmetry of causes must be found in effects; 
when some effects reveal some asymmetry, that asymmetry must be found in causes.}" Or in a  
more cursive way: {``{\sl C'est la dissym\'etrie qui cr\'ee le ph\'enom\`ene}"}. \\
An example is provided by the phenomenon of piezoelectricity, \ie the creation of an electric (vector) field 
$E$  in a crystalline material subject to a mechanical stress. The latter is described by a rank-two tensor $u$,   
in a linear approximation the electric field is proportional to $u$, 
$E_i=\sum_{jk} \gamma_{i,jk} u_{jk}$,
and hence the phenomenon depends on the existence of a non vanishing rank 3 tensor 
$\gamma_{i,jk}\ne 0$; 
 if the crystal admits a symmetry  by ``inversion", (\ie reflection with respect to a 
point), $\gamma_{i,jk}$ is changed into $-\gamma_{i,jk}$ under inversion and must vanish, 
and this rules out piezoelectricity in many crystal classes. 
Only non symmetric crystalline classes may give rise to piezoelectricity. 
\\
Curie also understood that breaking of a symmetry under a group $G$ may leave invariance
under a subgroup (an ``intergroupe" in his terms) $H$ of $G$, an idea still quite topical. 
For instance he classified the possible breakings and subgroups 
of a system invariant under rotations around an axis, \ie under the group $D_\infty$ in modern terminology.

\smallskip
{\bf Limits of group theory.}  {As noticed by M. Senechal \cite{Senechal}, 
``group theory cannot answer a question that seems fundamental today: which shapes
tile space and in what way?". That question has of course become highly relevant since the discovery some 30 years
ago of quasicrystals. In this new class of materials, rotational order does not extend to large distances
and translation invariance is lost. Still diffraction of X-rays  leads to patterns of bright spots exhibiting some symmetry.
This has led the International Union of Crystallography to redefine the term ``crystal" so as to include both ordinary periodic crystals and quasicrystals. According to this new definition, a crystal is
 ``any solid having an essentially discrete diffraction diagram".}


\vglue-6mm
\section{{Special Relativity and Lorentz group: Lorentz, Poincar\'e, Einstein 
\dots}}

Special Relativity is often regarded as the first appearance of Lie group theory in modern physics. Let us recall
some of the crucial steps, referring the reader to more scholarly sources \cite{Pais, Darrigol} for further details.

\bei
\item{Lorentz} (1892-1904) (after Voigt and FitzGerald, and in parallel to Larmor)
discovers what are now called the Lorentz transformations 
and the resulting contraction of lengths. 
His purpose is to make the Michelson-Morley experiment consistent with the existence of aether.
\item {Poincar\'e} (1905) establishes the covariance of Maxwell equations under Lorentz transformations; 
he also sees that Lorentz transformations together with space rotations
leave the form {$x\cdot x:=x^2+y^2+z^2-c^2t^2$} 
invariant and form a group, thus giving them their proper geometric meaning, much in the spirit of Klein.
In his approach, however, the Lorentz group is not derived from first principles. 
\item {Einstein} (1905) starting from two principles -- (i) 
 the principle of relativity: physical
laws do not depend on the inertial frame of the observer; and (ii) in an inertial frame
the speed of  light $c$ is an absolute  constant of Physics,  
independent of the uniform motion of the source -- {\it constructs} the Lorentz transformations; he notices as a 
side remark that Lorentz  special transformations (or ``boosts" as we call them now) 
of {\it colinear} velocities form a group, {``wie die sein muss"} (as they should)\footnote{the only occurrence of the word ``Gruppe" in his paper\dots}; he proves that they leave  Maxwell's equations invariant,
but does not seem to notice or at least does not comment on the fact that they also preserve
the form $x\cdot x$. 
\item {Minkowski} (1908) introduces ``space-time", identifies the Lorentz group as the invariance group of the metric 
{$x_1^2+x_2^2+x_3^2+x_4^2$} with $x_4=ict$,  makes use of the notion of 4-vectors and tensors, and  shows 
 the covariant way of writing Maxwell equations.  At first Einstein is not impressed by this piece of work,  
qualifying it as {``\"uberfl\"ussige Gelehrsamkeit"} (superfluous erudition)! \cite{Einstein-AP}.
After starting to work on gravitation, however,  Einstein soon realizes the power of tensor methods\dots
\end{itemize} 

Thus, although Einstein made a real breakthrough in physics and utterly changed  our view of space and time
by ``propounding a new chronogeometry" \cite{Darrigol}, it seems fair to say that group theory played a 
very minor role in his work and his lines of thought.

More mathematically inclined people thought otherwise. We have already mentioned Poincar\'e's and
Minkowski's works. {Klein} (1910) \cite{Klein} 
observes: 
{``One could replace `theory of invariants relative to a group of transformations' by the words
`relativity theory with respect to a group'."} For him, Galilean  relativity or Special Relativity were clearly in the
straight line of his Program.

\section{General Relativity\dots\ and gauge theories}

General Relativity is an emblematic case   illustrating Klein's program in 
a differential geometric  context. There, following Einstein's vision,  one postulates the 
 invariance of the equations of the gravitational field under general coordinate transformations. And by a sort of
 reverse engineering, one looks for equations knowing the invariance group. This is what was
 achieved by {Einstein and by Hilbert} (1915), with the celebrated equation 
 \be\label{E-H} R_{\mu\,\nu}-\oh R\,  g_{\mu\,\nu}= \kappa T_{\mu\,\nu}\,,\ee
  with $R_{\mu\nu}$ the Ricci tensor, $R=R^\mu_\mu$ its curvature,  $T_{\mu\nu}$ the energy-momentum tensor, and 
  $\kappa =  \frac{8\pi G}{c^4}$ where $G$ is Newton's gravitational constant.
 Recall that Hilbert derived this equation from the invariant action
    \be S = \int \left[ {1 \over 2\kappa} \, R + \mathcal{L}_\mathrm{M} \right] \sqrt{-g}\,d^4x\,, \ee
 with ${\cal L_\mathrm{M}}$ describing the invariant coupling of gravity to matter, and $T_{\mu\nu}=\partial ({\cal L}_\mathrm{M}  \sqrt{-g})/\partial g^{\mu\nu}$.
 I shall not dwell more on that subject, as it was treated by other contributors to the meeting. 

\def\Gd{\delta}
 Let me rather make a big leap forward in time, and observe that a similar approach was taken in the construction
 of non-abelian gauge theories. The gauge invariance of electrodynamics had been observed by Weyl (1918) and 
 reformulated  later by him into what we now call   U(1) gauge invariance \cite{Weyl18, Weyl29}. 
 Looking for a generalization to non abelian groups $G$, \ie postulating invariance under a certain {\it infinite dimensional} group
 of local, space-time dependent transformations,
 Yang and Mills (1954) \cite{YangMills54} were led to an (essentially) unique solution, 
 with Lagrangian density
 \be {\cal L}=\frac{1}{2g^2} \tr F_{\mu\nu}F^{\mu\nu}, \qquad F_{\mu\nu}=\partial_\mu A_\nu -\partial_\nu A_\mu-[A_\mu,A_\nu]\,,\ee
 with  the gauge field $A_\mu$ and its field strength tensor $F_{\mu\nu}$ (a connection and its curvature on a fiber
 bundle) taking values in the 
 Lie algebra of $G$ or one of its representations.  Here and below, $\partial_\mu$ stands for 
 $\frac{\partial}{\partial x^\mu}$. ${\cal L}$ is invariant under local infinitesimal changes 
 $\delta A_\mu(x)= D_\mu \Gd \alpha(x)$, with $D_\mu=\partial_\mu - [A_\mu,\cdot]$ the covariant derivative, and $\Gd \alpha\in 
 {\rm Lie}\,G$. A term ${\cal L}_\mathrm{m}$ may then be added to ${\cal L}$  to describe the gauge invariant coupling to matter. 
 This now famous and ubiquitous  Yang-Mills theory is  the cornerstone of the Standard Model of particle physics, see below.
 
 To summarize, here are two cases  (Einstein--Hilbert, Yang--Mills) in which 
 invariances and geometry of space { (either real space-time or ``internal" space)} 
  prescribe the dynamics. According to Yang's motto \cite{Yang86},  
 ``symmetry dictates interaction".  

\vglue-8mm
\section{Emmy Noether: invariances and conservation laws}
Noether's celebrated paper (1918) \cite{Noether},  presented on the occasion of Klein's academic Jubilee, 
contains  two theorems on 
group invariance  in variational problems. I give a sketch of her results, using  modern terminology and
notations, and I refer to \cite{YKS} for a 
translation of her original article and a detailed and critical reading, see also \cite{Kastrup}.
\\
Consider a field theory described by an action principle  in a, say,  $4$-dimensional space-time with coordinates
$x=(\vec x,t\equiv x^0)$
$$S=\int {\cal L}(x\,;\ \phi^i(x),\partial \phi^i(x),\cdots) d^4 x$$
with $S$ the action and $\CL$ the Lagrangian density, a local function of a collection of fields   $\{\phi^i\}$ and of finitely many of their 
derivatives. Assume 
the invariance of ${\cal L}\,d^4 x$ (and hence of $S$) under a Lie group of
coordinate and field variations $x\mapsto x',\ \phi\mapsto \phi'$. Then 
{Noether's first theorem} asserts: 
\begin{theorem}[Noether]\label{Noether1} An $ n$ dimensional Lie group of invariance of  ${\cal L}\,d^4 x$ implies the existen\-ce of 
$ n$  independent divergenceless currents 
$$j^\mu_s=(j^0_s(\vec x,t), \vec j_s(\vec x,t))\,,\qquad \ie\quad 
\partial_\mu j^\mu_s{\equiv  \frac{\partial}{\partial t} 
j_s^0- {\rm div}\, \vec j_s}=0\,,  \quad s=1,\cdots, n\,,$$
 from which, by Stokes theorem, {$n$} independent conservation laws follow
$$  \frac{d}{dt}Q_s := \frac{d}{dt} \int d^{3} x\, j^0_s(\vec x,t) =\int d^{3} x\,\, {\rm div}\,\vec j_s(\vec x,t)=0 .$$\end{theorem}
(The currents are assumed to vanish fast enough at spatial infinity to justify the last step.)
Suppose that ${\cal L}$ depends only on $\phi$ and its first derivatives $\partial \phi$.
Write coordinate and field infinitesimal variations as
$ \delta x^\mu =X^\mu_s(x,\phi) {\delta a^s}$ and  $ \delta \phi^i= Z^i_s(x,\phi) 
{\delta a^s}$,  
 where $a^s,\  s=1,\cdots n$, are parameters in the Lie algebra, and 
 Einstein's convention of summation  over repeated indices is used. Then  one finds
\begin{align} 
j^\mu_s&=-\frac{\partial {\cal L}}{\partial \partial_\mu \phi^i}(Z^i_s -\partial_\rho \phi^i X_s^\rho) -X_s^\mu {\cal L}\\  
\partial_\mu j^\mu_s \, \,\delta a^s &= \sum_i \Psi_i\, \delta\phi^i {\rm\  \ where\  \ } \Psi_i:=\frac{\delta {\cal L}}{\delta \phi^i}
:=\frac{\partial {\cal L}}{\partial \phi^i}
-\partial_\mu \frac{\partial {\cal L}}{\partial \partial_\mu \phi^i}
\nonumber \\ &=0 {\rm\ \  by\ Euler-Lagrange\ equations}\,. \nonumber
\end{align}
In Noether's paper, the converse property, namely that conservation laws imply invariance, is also derived.
This first theorem was subsequently generalized by Bessel-Hagen (1921) to the case where ${\cal L}\,d^4 x$ 
is invariant up to a total divergence   $  \delta a^s\, \partial_\mu  k_s^\mu\, d^4 x$, in which case $j_s^\mu$ is just 
modified by the additional  term  $ k_s^\mu$.

  As an example, consider a theory involving a complex scalar field $\phi$ with Lagrangian 
  $\CL = \partial_\mu \phi^* \partial^\mu \phi -V(\phi^* \phi)$, $V$ some arbitrary polynomial potential.
  The Lagrangian is invariant under the U(1) group of transformations $\phi(x)\to  e^{i\alpha}\phi(x)$, leading to
  a conserved Noether current $j_\mu(x)= i(\phi^*(x) \partial_\mu \phi(x) -(\partial_\mu\phi(x))^*  \phi(x))$. The 
  associated conserved U(1) charge may be  thought of as an electric (or baryonic or leptonic\dots) charge. 

Thus Noether's first theorem establishes a link between invariances under continuous transformations and conservation laws.
This was not a new result in physics. There had been 
 early precursors: {Lagrange} (1811),  {Hamilton} (1834), {Jacobi} (1837) had uncovered the fundamental
 conservation laws of energy, momentum and angular momentum in classical mechanics, but did not make a 
 systematic connection with geometric invariances. This had been elaborated by Sch\"utz (1897) and by other 
 precursors of  Noether:
 Hamel (1904) who introduced the calculus of variations in that context,  {Herglotz} (1911),  Engel (1916) 
 and Kneser (1917)
 who applied it to the 10 conservation laws due to Galilean and to relativistic invariance, 
 see \cite{Kastrup, YKS}.   
 But  E. Noether  was the first to give a general and systematic derivation of  conservation laws, 
starting from invariance of an action principle under Lie algebraic transformations. 

This important result of Noether had a curious fate. 
After initial applause by Klein, Hilbert and others, and some generalization by Bessel-Hagen, 
 came a long freeze. That was caused mainly by the rise of
quantum mechanics, which  made no use of the Lagrangian formalism. Thus Noether's theorem was essentially
forgotten until the  early 50's, when  covariant Quantum Field Theory (QFT) developed, causing  a 
 revival of interest in the Lagrangian formalism, and Noether theorem became important again.\\
 
 In modern QFT, her theorem  appears in particular in the guise of {\it Ward--Takahashi identities}
  satisfied by 
 the vacuum expectation values of ``time-ordered products of fields", (T-products in short),
 which are the relevant Green functions.  
In the latter  the field operators are ordered from right to left according to increasing time,
  $T \phi_1(y_1)\cdots \phi_n(y_n) :=  \phi_{\pi_1}(y_{\pi_1})\cdots \phi_{\pi_ n}(y_{\pi_ n})$,
  with $\pi $ 
  a permu\-tation of $\{1,\cdots,n\}$ such that $y_{\pi_ 1}^0\ge y_{\pi_ 2}^0\ge \cdots\ge y_{\pi_ n}^0$.
   Take an ``internal" sy\-mme\-try ($X^\mu_s  =0$ in the above notations), consider 
   its Noether   currents $ j_s^\mu$ and the divergence of its time-ordered product with fields
   $\langle  T j_s^\mu(x) \phi_1(y_1)\cdots \phi_n(y_n)\rangle$. In addition to the explicit divergence which vanishes
    because of the current conservation,  $\partial_\mu  j_s^\mu=0$,
   there is a contribution coming from the implicit Heaviside functions $\theta(\pm(x^0-y_{i_p}^0))$ in the T-product.
   Then one finds
   \begin{align}\nonumber  \frac{\partial}{\partial x^\mu} 
 \langle T j_s^\mu(x) \phi_1(y_1)&\cdots \phi_n(y_n)\rangle  \\
  \label{Ward}  &=\! \sum_{i=1}^n
 \delta(x^0-y_i^0)\,
 \langle T \phi_1(y_1)\cdots [j^0_s(x),\phi_i(y_i)] \cdots\rangle\end{align}
 and the equal time commutator   on the r.h.s.   is the density of the infinitesimal variation of 
 the field $\phi_j$:  
  $[j^0_s(x),\phi_i(y_i)]_{x^0=y_i^0}  = Z_s^i(x,\phi_i) \Gd^3(\vec x-\vec y_i)$.
  These identities lead to very useful relations between  different T-products. 
  
  In the case the symmetry is not exact but is ``softly broken" and one has a {\it partial conservation} of the 
  current $\partial_\mu  j_s^\mu(x) = \chi(x)$, with $\chi$ an explicitly known field, the content of the suitably modified 
 identity (\ref{Ward}) is not void but leads to relations between amplitudes that have been explored in great 
 detail, in particular in the context of  weak interactions. 
  
These identities and their various avatars -- in particular the Slavnov--Taylor and BRST (Becchi--Rouet--Stora--Tyutin)
identities in the framework of gauge theories -- play a 
crucial role at several steps of the study of quantum field theories. They enable one 
to  establish that the renormalization procedure does not jeopardize the symmetries of the 
original theory; they  allow one to prove that  conserved currents ``do not renormalize" and do not develop 
anomalous dimensions, thus justifying the  notion of universality in ``current-current" interactions, {see  below};
they are also used in the derivation of ``low energy theorems", see in particular 
\cite{Weinberg}.

For completeness, let us mention briefly Noether's {second theorem}:  for an ``infinite dimensional group" of invariance 
(such as diffeomorphisms in General Relativity, or gauge transformations in gauge theories), 
invariance within a variational principle implies the existence of constraints between the 
$\Psi_i=\delta{\mathcal L}/\delta\phi^i$, i.e. identities satisfied independently of the Euler--Lagrange
equations of motion.  Examples are provided by the  contracted Bianchi identities in General Relativity,
$D^\mu G_{\mu\nu}=0$, where $G_{\mu\nu}=R_{\mu\nu}-\oh g_{\mu\nu}R$,  
 or their analogue $D^\mu D^\nu F_{\mu\nu}=0$ in gauge theories.  
Note that, although they are satisfied irrespective of the Euler--Lagrange equations, (\ref{E-H}) or $ \frac{1}{g^2}D^\nu F_{\mu\nu}= J_\mu
:= \partial {\cal L}_\mathrm{m}/\partial A^\mu$
respectively, these identities ensure the consistency of the latter, whose right hand sides are covariantly conserved
$D^\mu T_{\mu\nu}=0$, resp. $D^\mu J_\mu=0$. 
\ommit{\magenta This has the practical consequence that 
in such a theory, there are {\it less} conserved quantities than anticipated from Noether's first theorem.}\\


\section{Invariances in Quantum Mechanics}
With the triumph of Quantum Mechanics, 
 a new paradigm  appears in the study of symmetries in physical systems. 
 Through the fundamental papers and books of von Neumann  and
  Wigner,  Weyl and  van der Waerden \cite{vonN-Wigner28,Weyl28,Wigner31,vanderW32} at the end of the 1920's, 
 representation theory enters Physics. This is particularly well summarized in Wigner's theorem. With any
 quantum system is associated a Hilbert space ${\cal H}$. States of the system are described by 
 vectors {$\Psi$}, or more precisely by {\it rays}, of ${\cal H}$ and ``observables" {$A$} are self-adjoint
operators on  ${\cal H}$. Then  Wigner's theorem \cite{Weyl28, Wigner31} asserts
 
 \begin{theorem}[Wigner]Transformations of a quantum system under a group $G$  are implemented 
as {$\Psi\to U \Psi$, $A\to UAU^{-1}$} with $U$ unitary or anti-unitary and unique \textit{up to a phase}, satisfying 
  $$g,g'\in G\qquad {U(g)U(g')=U(g.g') e^{i\omega(g,g')}}\,.$$
Thus $U(g)$ gives a projective (up to a phase) representation of $G$.\end{theorem}

By ``anti-unitary", we mean a unitary antilinear operator, a situation which is encountered in the study of the time reversal
operator $T$.
Note that the projective nature of the representations is forced upon us by the structure of Quantum Mechanics: rays 
rather than vectors are the relevant objects.

\def\d{{\rm d}}
Among such transformations, 
invariances are associated with  group actions that commute with the dynamics, \ie with the Hamiltonian 
\be\label{comm-symm} [H,U(g)]=0\,.\ee
But according to Ehrenfest's theorem, the time derivative of any  operator (with no explicit time dependence)
is given by its commutator with $H$, $i\hbar\, \d A/\d t=[H,A]$.
Thus (\ref{comm-symm}) tells us that any $U(g)$ or any infinitesimal generator of the 
group action, is conserved:
here again, invariances manifest themselves by the existence of conserved quantities.
The new feature due to quantum mechanics is that not all conserved quantities
are simultaneously observable. If one picks  $H$ and a set 
of  commuting operators $U(h)$, 
($h$ in a Cartan torus of $G$ if $G$ is a Lie group), eigenstates of those $U(h)$ have conserved eigenvalues,  which are, 
 in the physicists' jargon,  ``good quantum numbers".

\ommit{Then as usual in such circumstances, one seeks a maximal set of commuting operators $H$ and $U(h)$, $h$ in a 
Cartan torus of $G$. 
Diagonalize a maximal commuting subgroup of $G$ :}

For example, consideration of the group of rotations SO(3) shows that its infinitesimal generators (\ie  elements of its
Lie algebra) are proportional to the components of the angular momentum $\vec J$. The latter is thus quantized 
by the theory of representations of SO(3). If the system under study is invariant under rotations, one has  
conservation of {$\vec J^2$} (the Casimir operator) and of one component, say {$J_z$}: their eigenvalues $j(j+1)\hbar^2$ 
and $m\hbar$ are ``good quantum numbers", conserved in the time evolution. 
States of the system are classified by  representations of {SO(3) or SU(2)}, the latter appearing because
it gives the projective (up to a sign) representations of the former, through half-integer spin representations. 

As a side remark, we also notice that the distinction between discrete and continuous invariances, 
that was crucial in classical physics, with only the latter leading to conservation laws, fades away. 
Conservation of parity -- to the extent it is conserved -- is expressed by the commutation relation 
$[P,H]=0$ and implies that  the parity of a state is a good quantum  number.

This beautiful framework was 
first applied to the rotation group and its finite subgroups, in conjunction with parity and the symmetric group of permutations.
The latter appears in connection with the Pauli
principle and the Fermi-Dirac or Bose-Einstein quantum  statistics. This resulted
in  innumerable applications to atomic, molecular and solid state physics:
atomic and molecular orbitals, 
the fine structure of spectral lines of atoms and their splitting  in a magnetic or electric field (respectiveley the Zeeman and
the Stark effects), the crystal-field splitting and many other effects were analysed by  group theoretic methods; 
 selection rules in transitions were shown to be governed by tensor products of representations, etc. 
 See for example \cite{Tinkham} for a review, and \cite{Scholz} for an overall  presentation of the work 
 of the first actors -- Wigner and von Neumann,  Heitler and London, Weyl.
First applications to particle physics were exploiting rotation, parity  
and Lorentz invariance in scattering theory.  In the latter context, let us 
cite Wigner's fundamental work on the representations of the Poincar\'e group \cite{Wigner39}. For a one-particle state, 
these representations are fully characterized by two real numbers, which describe the mass and
the spin of the particle. But more group theory was soon to come in particle physics
and we devote the next section to these new symmetries.

As it is often the case when a new theoretical corpus develops, requiring the learning and the practice  
of an abstract formalism, not everybody accepted happily this irruption of group theory into physics
and  there was a certain resistance among some physicists.  
Some even talked about
{``the group pest"}!\dots, see  \cite{Wigner81},  \cite{vanderW32} p. 165, \cite{Scholz}, or the prefaces of \cite{Weyl28, Sternberg}.
In his preface to the 1959 edition of his book \cite{Wigner31}, Wigner observes: ``It pleases the author that this
reluctance [among physicists toward accepting group theoretical arguments] has virtually vanished in the meantime and that, in fact, the 
younger generation does not understand the causes and the basis for this reluctance."


\section{Invariances in particle physics}

We have seen above that  Noether's reciprocal statement enables one to 
infer the existence of a symmetry group from  conserved quantities. This 
observation  has been beautifully illustrated by the discovery of
 ``flavor groups" in particle physics.\\
 {Heisenberg} (1932) observing the many similarities of mass and interactions of the two constituents
 of the nucleus, the {\it  nucleons}, namely 
 the neutron $n$ and the proton $p$, their electric charge notwithstanding, 
 proposed that they form a 2-dimensional representation of a new {SU(2)} group of ``isotopic spin",
 or ``isospin" in short. 
 This was an extremely fruitful idea, soon confirmed by the discovery (1947) 
 of the $\pi$ mesons, or pions, coming in three states   of charge $(\pi^+,\pi^0,\pi^-)$}, 
 and hence forming a 3-dimensional representation of this SU(2) group. Isospin symmetry then predicts 
relations between scattering amplitudes of nucleons and pions that were well
 verified in experiments.  
 Later, more instances came with the {kaons $(K^+,K^0)$},  the $\Delta$ resonance
 {$(\Delta^{++},\Delta^+,\Delta^0,\Delta^-)$} and others, which
  form  representations of isospin $1/2$, $3/2$ \dots\ respectively. This SU(2) group is a
 symmetry of {\it hadrons} (\ie of strongly interacting particles), broken by electromagnetic interactions.
 
 In the sixties, the story repeated itself. In view of the newly discovered ``strange" particles, 
{Gell-Mann and Ne'eman} (1961)  
proposed the existence of an SU(3)  group of (approximate) symmetry of strong interactions.
This  ``flavor SU(3)" group encompasses the previous isospin group SU(2).
The argument leading to SU(3) 
was that   there was experimental evidence of the existence of
 {\it two} independent conserved quantities (isospin and {\it hypercharge} or {\it strangeness}),
hence the group should be of rank 2. Also there were several observed ``octets" (8-dimensional representations) of
particles of similar masses and same quantum numbers (baryonic charge, spin,  parity), and this pointed to the 
group SU(3) which has an 8-dimensional irreducible representation, namely its adjoint representation. 
This hypothesis was confirmed soon after by the experimental discovery of a particle $\Omega$ completing a 
10-dimensional representation, whose  mass
and quantum numbers had been predicted, and by some other experimental evidence \cite{Gell-Mann-Neeman}.
Associated with the fundamental 3-dimensional representation of SU(3) is a triplet of ``quarks", 
$(u,d,s)$ (for up, down and strange), which according to the confinement hypothesis, should not
appear as observable particles in normal circumstances\footnote{By ``normal circumstances"
we mean discarding the extreme conditions of the primordial Universe, immediately after the 
Big Bang, or of the high-energy heavy ion collisions in the laboratory, 
where a plasma of unconfined quarks may be created. }. 
The  SU(3) group has been dubbed ``flavor" to distinguish it from 
another ``color" SU(3) that appears as the gauge group of ``quantum chromo-dynamics" (QCD), the modern theory
of strong interactions. 
To conclude this discussion, let us stress that the flavor SU(3) group of (approximate) 
symmetry was more than welcome, in order to put some order and structure
in the ``zoo" of particles that started to proliferate at the end of the fifties.

This line of thought  has proved extremely fruitful, and 
modern particle physics has seen a blossoming of discoveries 
structured by the concepts of symmetries and group theory. 
The previous SU(2) and SU(3) groups have been extended to larger flavor groups, in 
connection with the discovery of new families of particles, with new quantum numbers, 
revealing the existence of  more species (or ``families") of quarks. 

The role of symmetries is not 
limited to strong interactions and the other subatomic forces -- electromagnetic and  weak --
are also subject to symmetry requirements. This was hardly apparent in the early  Fermi ``current-current"
theory of the weak interactions, ${\cal L}_{\rm F}= -\frac{G_F}{\sqrt{2}} J^\mu J_\mu$, 
but then  the V-A pattern  \`a la
Gell-Mann--Feynman of the current $J=V-A$, the role of the conservation or partial conservation of
currents $V$ and $A$, the Cabibbo angle, etc. were gradually uncovered,
see \cite{PaisIB, Ilio} for reviews of these historical developments. This role of symmetries
is even more  manifest nowadays in 
the Glashow--Salam--Weinberg model of electroweak interactions, see below. 

To look for a  group invariance whenever a new  pattern is observed
has become a second nature for particle physicists. 


\section{The many implementations of symmetries in the quantum world}

When discussing symmetries in contemporary  physics, it is common to distinguish 
space-time symmetries, discrete or continuous,   
-- rotations and Lorentz transformations, translations, space or time reflections, \dots --
from ``internal" symmetries that act on internal degrees of freedom -- charge, isospin, etc. 
While this distinction may be useful, it should not hide the tight interlacing of these two species
of symmetries. For instance, one of the fundamental results in QFT is the CPT theorem (L\"uders, Pauli and Bell) 
which asserts that the product of the  charge conjugation $C$ by the space reflection or parity $P$
and time reversal $T$ should be an absolute and uninfringed symmetry of Nature.  This is 
established based on fundamental properties like locality and Lorentz invariance 
 that one expects from any decent theory \cite{StreaterWightman}\footnote{As pointed out by Yang   \cite{Yang86}, there is a very intriguing sentence in Weyl's preface to the second edition of his book \cite{Weyl28},
   which seems to indicate that as early as 1930, he  foresaw 
 some relation between these three transformations.}.

Another distinction between two big classes of symmetries deals with their ``global" or 
``local" character. The isospin SU(2) or the flavor SU(3) symmetries mentioned above are global symmetries,
in the sense that the group element describing the transformation is independent of the space-time 
point where it applies. In contrast, the diffeomorphisms of GR or the gauge transformations of
electrodynamics or Yang--Mills theory are local, with the group (or in infinitesimal form, the Lie algebra)
element varying from point to point. As we have seen, that distinction was already clearly perceived by Klein 
and Noether.

 \medskip

It turns out that 
a {quantum symmetry} may be realized in a multiplicity of ways, namely
 \begin{itemize}
 \item as an {\it exact} symmetry, {\it e.g.} in the global U(1) symmetries associated with charge or baryonic number
 conservation, or in the local gauge invariances of  quantum electrodynamics and of quantum chromodynamics (QED, QCD);
 \item as an explicitly broken symmetry: this is the case with isospin SU(2) broken by electromagnetism, 
 or flavor SU(3), which is an approximate symmetry, broken by the strong interactions themselves. This is also 
 the case with parity, the space reflection $P$ mentioned above, which is explicitly broken by weak interactions,
 as discovered by Lee and Yang (1956) and as now implemented  in the Standard Model;
 \item as a  spontaneously broken symmetry. This refers to the following situation:  
 in a physical system {\it a priori} endowed with a certain symmetry, the state of
 minimum energy, called the  ground state or the vacuum depending on the context,  may in fact be {\it non 
 invariant}.
 This is a very common and fundamental phenomenon, which 
 is familiar from the case of ferromagnetism: in a  ferromagnet in its low-temperature phase,  
 the magnetic moments of the individual atoms,
 although subject to a rotation invariant interaction, pick collectively a direction in which they align 
 on average, thus giving rise to a macrocopic magnetization that breaks the rotation invariance of the 
 whole system. 
 This is accompanied, if the broken symmetry is continuous, by the appearance of 
massless excitations or particles, associated with the possibility of continuously rotating the 
ground state at a vanishing cost in energy. These excitations are the {\it Nambu--Goldstone} 
particles. In the variant in which the symmetry is only approximate, 
and in the neighbourhood of a spontaneously broken phase,  one expects the would-be Nambu-Goldstone
 bosons to be not strictly massless but of low mass;
 \item as a  spontaneously broken  gauge symmetry: {a global symmetry is spontaneously broken 
but the resulting theory maintains an exact gauge invariance}. Then, and this is the essence of
 the Brout--Englert--Higgs (BEH) mechanism, the Nambu--Goldstone excitations do not appear as 
 real particles, and instead give rise to additional polarization states of some vector fields and to 
 masses of the corresponding particles. This is a crucial step in the edification of the electro-weak sector of the 
 Standard Model, and the successive discoveries of massive vector particles (the $W^\pm$ and $Z^0$)
 and lately, of a candidate for the relique massive scalar boson at CERN, seem to corroborate this model;
 \item anomalously, which means  through a breaking of a classical symmetry by quantum effects. 
 Examples are provided by the realization of some chiral symmetries of fermions, which 
 act separately on the 
 left-handed and right-handed components of these particles. Conversely, in the Standard Model of particle
 physics, 
 where the assignments of representations are different for different chiralities, it is essential
 that  anomalies cancel, see below;
 \item with supersymmetry: that ordinary Lie groups and algebras could be extended
 to accommodate anticommuting (Grassmannian) elements has been known and well studied since 
 the seventies. To this date we have 
 not seen any direct manifestation of supersymmetry in the laboratory. But the idea has been
 so amazingly fruitful in establishing new results and new connections between different
 fields that it will undoubtedly remain in the physicist's toolbox;
 \item as quantum symmetries, or ``quantum groups", a misnomer for ``quantum" deformations of
 Lie algebras or, more generally, for Hopf algebras. These have not yet manifested themselves in the context of particle 
 physics, but are determinant in the discussion of quantum integrable models and in their
 applications to many systems of  condensed matter physics in low dimension;
 \end{itemize}
 \dots and this list is certainly non exhaustive. \\
 It is truly remarkable  
 that Nature makes use of all these possible implementations of symmetries.\\

Let us illustrate these various possibilities on a few examples coming from modern physics.
Our presentation will be
extremely sketchy, as each topic would deserve a separate monograph\dots\\

{\bf Example 1}: ``{Linear/non linear sigma models}" may be regarded as 
Klein's most direct heirs in the context of QFT.\\
\ommit{Take a field {$\phi\in {\cal M}$}, a Riemannian manifold. Write a Lagrangian in the form
{$$ {\cal L} = (\partial \phi,\partial \phi) -V((\phi,\phi)) $$}
invariant under the isometry group of the metric. May impose further 
constraint $\forall x\quad (\phi(x),\phi(x))={\rm const.}$. For ex ``O(n) sigma model" on sphere $S^{n-1}$.}
In the simplest possible case, consider a field $\phi $ defined on $\mathbb{R}^d$ and taking its values  in $\mathbb{R}^n$ or {$\in S^{n-1}$} and write a Lagrangian in the form
{$$ {\cal L} =\oh (\partial \phi,\partial \phi) -V((\phi,\phi)) $$}
where $(\,,\,)$ denotes the  {O($n$)} invariant bilinear form. 
The invariance group of that Lagrangian is obviously {O($n$)}, and the field $\phi$ transforms according to a 
linear representation or to a non-linear realization, depending on the case $\RR^n$ resp. $S^{n-1}$.
 According to Noether's theorem, there are 
$\oh n(n-1)$ independent conserved quantities at the classical level. Using the corresponding Ward identities (\ref{Ward}),
one verifies that the symmetry is preserved by quantum corrections.
This was first set up by {Gell-Mann and L\'evy} (1960) in the case $n=4$,  in their  investigation of  the partial conservation of the 
``axial current" in weak intercations\,\cite{GML}, and involved 
the fields of pion particles $\pi^\pm, \pi^0$ and of a hypothetical field $\sigma$, 
whence the name given to the model; this original model had thus a (softly and spontaneously broken)  O(4) symmetry.

This may be generalized to  a field $\phi$ taking its values in ${\cal M}$, a Riemannian manifold with isometries. 
\ommit{Particularly interesting is the case of symmetric 
spaces $G/K$, which find applications ranging from condensed matter to cosmology\dots \\}
Now in any of these sigma models, 
 the natural questions to ask are \begin{itemize} 
\item{} how is the symmetry realized, as an exact, explicitly broken, or spontaneously broken symmetry?
\item{} how is the symmetry preserved by renormalization?  This is where use has to be made of  Noether currents and Ward identities;
\item{} what are the physical consequences: are there Goldstone particles, or  ``almost Goldstone" particles (like the pion of
 low mass)? is there a dynamical generation of mass?  is the theory scale or conformally invariant? and so on, and so forth.
\end{itemize}

{Sigma models have been extensively used with  all kinds of manifolds and groups in particle physics and cosmology, 
in statistical mechanics
and solid state physics. For example they appear as effective low-energy theories for various phenomena in condensed matter,
describing membranes, surface excitations, order parameters, \dots;  
but also in string theory -- again in  a low energy limit --, based on ordinary manifolds or  generalized geometries \`a la Hitchin. 
  The study of non compact and/or supersymmetric sigma models is a currently very active subject, 
  for its applications running from condensed matter to string theory. 
  
These sigma models also constitute a mine of mathematical problems. For instance, 
particular cases  with $V=0$ are  studied for their own sake, in Riemannian geometry, 
under the name ``harmonic maps".

\medskip
{\bf Example 2}: {The Standard Model of particle physics} has a symmetry group {SU(3)$\times$SU(2)$\times$U(1)},
with three gauge groups realized in a completely different way.

The SU(3) color (gauge) symmetry of QCD is an {exact} invariance, and  this is believed to be of crucial importance 
for quark confinement.
On the other hand SU(2)$\times$ U(1), the gauge group of 
weak isospin and weak hypercharge, is {spontaneously  broken} down to an {\it exact} {U(1)}, the gauge symmetry  of 
ordinary electrodynamics. As mentioned above, a relique of the BEH mechanism at work in this
spontaneous breaking should be  a spin 0 boson, a good candidate of which has just   been observed at CERN.

The absence of anomalies in the Standard Model, crucial for the consistency of the theory,  relies on a remarkable
matching between families of leptons and of quarks:
for both types of particles three ``generations" are known at this time 
$$(e,\nu_e),\ (\mu,\nu_\mu),\ (\tau,\nu_\tau) \longleftrightarrow {(u,d),\ (c,s),\ (t,b)}\,,$$
and anomalies cancel within each generation \cite{cancel-anomalies}.
 
 On top of the gauge pattern, there are {\it other} SU(2) and SU(3)  groups at work:
the flavor {SU(2)$\subset$ SU(3)}  broken  symmetries discussed above. 
In another vein, 
a scenario which has been contemplated -- and in fact studied in great detail --  but does not  yet seem 
to be borne out by experiments 
is that this Standard Model is in fact a subsector of a larger supersymmetric extension. 
\medskip

{\bf Example 3}: {Quantum integrable systems and Quantum Groups}\\
Consider the { spin $\oh$} XXZ quantum chain: this is a quantum system of $N$ spins whose
interactions  are described by the Hamiltonian%
{$$ H=\sum_{i=1}^N  S_i^xS_{i+1}^x+S_i^yS_{i+1}^y+ {\Delta}\, S_i^zS_{i+1}^z+{\rm boundary\ terms}$$}
acting in  $(\mathbb{C}^2)^{\otimes N}$. $\Delta$ is an anisotropy parameter in spin space. It was first introduced for 
$\Delta=1$ by  Heisenberg (1928) as a model of ferromagnetism.  This   is known to be 
a quantum  integrable system after important contributions by {Bethe, Lieb and Sutherland, Yang and Yang, Gaudin, Baxter, 
Faddeev and many others}.
For $\Delta=1$, (and no boundary term), it exhibits SU(2) invariance. 
For $\Delta\ne 1$, $|\Delta|<1$, it has a deformed symmetry $U_q sl(2)$ (``quantum SU(2)"), 
where {$q=e^{i\alpha}$, $\Delta=\cos \alpha$} \cite{Pasquier-Saleur}, or an affine quantum $U_q \widehat{sl}(2)$}
\cite{Miwa}, depending on the boundary conditions. 
Recent progress on the computation of correlation functions of the XXZ chain and
on its connections with problems of combinatorics
have been made possible by representation theoretic
considerations. 

Other recent advances  in the context of integrable gauge theories
and the AdS/CFT correspondence also rely to a large extent on representation theory  of quantum algebras.\\

{\bf Example 4:} Conformal invariance. The last fifteen years of the previous century have witnessed 
rapid  progress in our understanding of  quantum field theories in low dimension.
In 2 d, conformal invariant field theories (CFTs) have 
experienced a spectacular development, with a huge number of exact  results
and applications to critical phenomena and to string theories,  thus
writing  a new chapter of non-perturbative quantum field theory.
For the largest part, this progress was made possible by advances at the end of the seventies in
the  representation theory of infinite dimensional Lie algebras -- Virasoro, affine Lie algebras and their cousins --
that are the relevant symmetries of CFTs. For a review, see for example \cite{DFMS}.
There one sees once again the close ties between symmetries, group theory and their physical implications.

\newpage
 
\section{Conclusions}

We have seen that symmetries and group theory play an {essential} role in modern physics.
Their role  is:\\
-- to {\it dictate} the possible form of interactions on geometrical grounds: the cases of General
Relativity or of gauge theories are exemplary in that respect; but one may also quote  non-linear sigma-models, in which the form
of the Lagrangian is prescribed by the geometry of the manifold and the  isometries are playing a key role;\\
-- to {\it predict}: more invariance means  less independence, implying relations between different phenomena,
selection rules, {\it a priori} determination of multiplicities, etc., as illustrated by scores of examples in atomic, molecular,  
solid state and particle physics;  
and to {\it organize} a wealth of data, of particles, of phenomena: we have seen that representation theory
is instrumental in this {undertaking}; \\ 
-- to {\it protect} in the quantization (and renormalization). Once again, take the example of a
 gauge theory. Were its symmetries 
 broken by quantum effects (ultraviolet divergences, anomalies), the theory would lose
most of its predictive power or even become inconsistent. So we have a self-consistent picture, where symmetry implies constraints
(in the form of Ward identities), that in turn guarantee that symmetry is preserved by quantization.
This scheme is implemented recursively in the perturbative construction of gauge theories. 

\ommit{\magenta Conversely, in the absence of symmetry, there is no protection,  and  ``mixings" become possible. 
Example: mixing generations
of quarks or leptons, which are not controlled by symmetry arguments. }

\medskip
The study of  groups and  of representation theory is now part of the education of a modern physicist.
Some domains of representation theory -- of superalgebras, of quantum groups and of infinite
dimensional algebras -- have developed recently thanks to the incentive of physical applications. 

\medskip
Could a unified theory based on geometry and embracing all fundamental interactions including gravitation 
be constructed? That was Einstein's dream, this is still regarded today as the Holy Grail by many people, 
string theorists among 
others.

\medskip
We have emphasized the many possible implementations of symmetries in (quantum) physics. 
We have also  stressed that not only the nature of 
the symmetry group but also  the scheme of its breaking, and the residual subgroup of symmetry, are
determinant. In that respect, we are still living in the legacy of Klein and Curie\dots

\newpage
\section*{Acknowledgements} It is a pleasure to thank the organizers, 
Lizhen Ji and Athanase Papadopoulos, for inviting me to this meeting, and Michel Bauer,
Matteo Cacciari, Robert Coquereaux,   Yvette Kosmann-Schwarzbach and Raymond Stora
for several useful suggestions on a first draft of the {manuscript}.

\end{document}